\documentclass{svjour3}                     
\smartqed  
\usepackage{graphicx}
\journalname{Few-Body Systems (FB20)}

\begin{document}

\title{The universality of the Efimov three-body parameter
\thanks{Presented at the 20th International IUPAP Conference on Few-Body Problems in Physics, 20 - 25 August, 2012, Fukuoka, Japan}
}


\author{J. P. D'Incao \and J. Wang \and B. D. Esry  \and C. H. Greene
}


\institute{J. P. D'Incao \at
              Department of Physics and JILA, University of Colorado, Boulder, CO 80309, USA \\
              \email{jpdincao@jila.colorado.edu}           
              \and
              J. Wang \at
              Department of Physics, University of Connecticut, Storrs, CT 06269, USA \\
              \email{jiawanghome@gmail.com}
              \and
              B. D. Esry \at
              Department of Physics, Kansas State University, Manhattan, Kansas 66506, USA \\
              \email{esry@phys.ksu.edu}
              \and
              Chris H. Greene \at
              Department of Physics, Purdue University, West Lafayette, IN 47907, USA \\
              \email{chgreene@purdue.edu}
}

\date{Received: date / Accepted: date}

\maketitle

\begin{abstract}
In this paper we discuss the recent discovery of the universality of
the three-body parameter (3BP) from Efimov physics. This new result was identified 
by recent experimental observations in ultracold quantum gases where the value of the $s$-wave scattering length, $a=a_-$, at which
the first Efimov resonance is created was found to be nearly the same for a range of atomic species --- if scaled as $a_-/r_{\rm vdW}$, 
where $r_{\rm vdW}$ is the van der Waals length.
Here, we discuss some of the physical principles related to these observations that emerge from
solving the three-body problem with van der Waals interactions in the hyperspherical formalism. We also
demonstrate the strong three-body multichannel nature of the problem and the importance of properly accounting for nonadiabatic effects.
\end{abstract}

\section{Introduction}

In the past few years the field of ultracold quantum gases has
provided one of the best scenarios for exploration of one of the most intriguing quantum phenomena present in a system of just a few particles, namely, the 
Efimov effect \cite{S_Efimov}. In addition to providing the first experimental observation of this effect 
\cite{S_133Cs_IBK_1}-\cite{S_85Rb_JILA},
experiments have now shown that one of its most fundamental assumptions must be reconsidered. In his original
work, Efimov derived the form of the three-body effective interaction that needed to be
``regularized" at short distances to prevent the divergence of the ground-state energy (known as the Thomas collapse \cite{Thomas}).
This regularization, according to some simple and reasonable arguments, should be arbitrary, implying that
the energy of the Efimov states, as well as any other scattering observable, should depend on a single, system dependent, 
three-body parameter (3BP) incorporating all the details of the interactions at short distances. This assumption, however,
has been proven to be untrue for ultracold atoms. Ultracold quantum gas experiments \cite{S_133Cs_IBK_1}-\cite{S_85Rb_JILA} have shown that 
the value of the $s$-wave scattering length at which the first Efimov state appears, which we denote by the value $a=a_{-}$, is between $-8 r_{\rm vdW}$ and $-10 r_{\rm vdW}$, 
where under the previous assumption one would expect its value to be randomly located at values larger than $ r_{\rm vdW}$. [Here, $r_{\rm vdW}=(m C_{6}/\hbar^2)^{1/4}/2$ is the van der Waals length, 
characterizing the range of the typical $-C_{6}/r^6$ interaction between two neutral particles of mass $m$.]

This astounding experimental observation has sparked a great deal of theoretical interests, 
and the physical mechanisms behind this observation is still under debate. Although very interesting theoretical models have been 
explored \cite{S_Ueda_He}-\cite{S_Ueda}, we will focus here on the one we recently developed \cite{S_3BP_BBB} 
using the adiabatic hyperspherical representation. Specifically, in this paper we elaborate our physical picture through which the universality of the 3BP can be understood,
and we also demonstrate the strong three-body multichannel nature of the problem.

\section{Brief theoretical background}
 
Our analysis of the universality of the three-body parameter is based on the adiabatic hyperspherical 
representation. The following is a sketch of its basic aspects and fundamental equations. A more detailed description
can be found, for instance, in Refs.~\cite{S_DIncaoJPB,S_SunoHyper}. We start here from 
the hyperradial Schr{\"o}dinger equation:
\begin{eqnarray}
\left[-\frac{\hbar^2}{2\mu}\frac{d^2}{dR^2}+W_{\nu}(R)\right]F_{\nu}(R)
  +\sum_{\nu'\neq\nu} W_{\nu\nu'}(R) F_{\nu'}(R)=E
  F_\nu(R).\label{Sradeq}
\end{eqnarray}
The hyperradius $R$ can be 
expressed in terms of the interparticle distances for a system of three equal masses $m$,  
$R=3^{-1/4}(r_{12}^2+ r_{23}^2+r_{31}^2)^{1/2}$, and it is a coordinate that describes the overall size of the system; 
$\nu$ is a collective index that represents all quantum numbers necessary to
label each channel; $\mu=m/\sqrt{3}$ is the three-body reduced mass for three identical bosons;
$E$ is the total energy; and $F_{\nu}$ is the hyperradial wave
function. The simple picture resulting from this representation originates
from the fact that nonadiabatic couplings $W_{\nu\nu'}$ are the quantities
that drive inelastic transitions, and from the fact that the effective hyperradial potentials 
$W_{\nu}$, like any usual potential, support bound and resonant states, but now for
the three (or more) particle system.  

For our present study of three identical bosons in the total angular momentum state 
$J^{\pi}=0^+$ ($\pi$ being the parity), the Lennard-Jones potential is adopted to model the interatomic interactions:
\begin{eqnarray}
v^{a}_{\lambda}(r) = -\frac{C_{6}}{r^6}\left(1-\frac{\lambda^6}{r^6}\right),
\end{eqnarray}
where $\lambda$ is a parameter adjusted to give the desired $a$ and number of bound states. For convenience,
we denote the values of $\lambda$ at which there exist zero-energy bound states ($|a|\rightarrow\infty$) as 
$\lambda^*_n$, where $n$ corresponds to the number of $s$-wave bound  states. As the number of bound states increases,
however, the solutions of Eq.~(\ref{Sradeq}) become more complicated and a specialized technique \cite{S_JiaSVD} has been developed to 
treat such cases.

\section{Universality of the three-body parameter}

\begin{figure}[htbp]
\includegraphics[width=0.9\columnwidth,angle=0,clip=true]{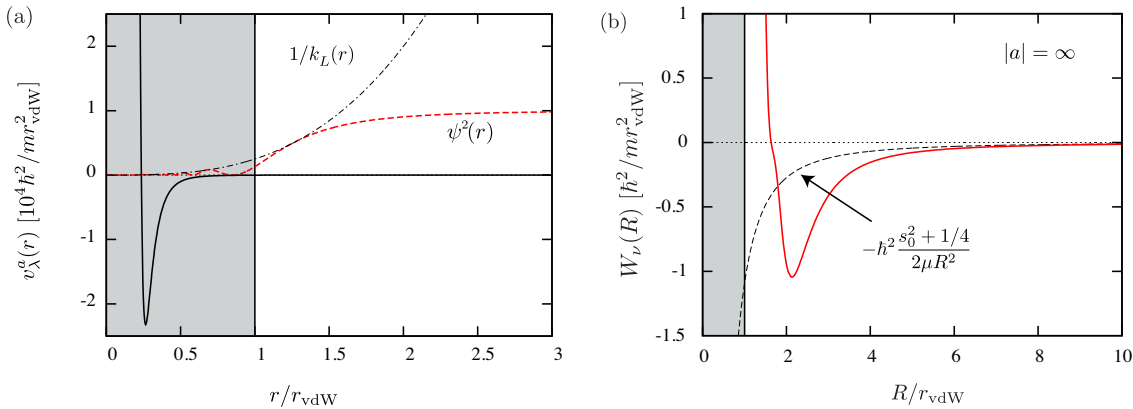}
\caption{(a) Evidence of the suppression of the two-body wavefunction due to the increase of the classical local velocity $k_{L}(r)$.
Here, the solid-black curve is the potential $v_{\lambda}^a(r)$ with $\lambda=\lambda^*_{10}$, the red-dashed curve is the numerical solution
for $\psi^2(r)$ and the black-dot-dashed curve is the suppression
factor $1/k_{L}(r)$. (b) Corresponding three-body effective potential
for the model in (a) with $|a|=\infty$. This potential (red-solid curve) is the one relevant for Efimov physics 
where $s_{0}\approx1.00624$ and it clearly displays the repulsive barrier for $R<2r_{\rm vdW}$ which suppresses the probability 
of finding particles at shorter distances and leads to the approximate universality of the 3BP.}
\label{Fig1}
\end{figure}

Reference~\cite{S_3BP_BBB} solves the three-body problem using a single channel model for the interatomic interaction
and shows that, when it is described by a Lennard-Jones potential (which has the proper $-C_{6}/r^6$ asymptotic behavior),
three-body observables are largely independent of the details of the interactions, i.e., of the number of
two-body bound states as well as other truly short-range three-body forces. In our physical picture, it is the sharp cliff
due to the $-C_{6}/r^6$ two-body interaction that leads to the universality of the 3BP. In this case, 
when particles approaches distances $r<r_{\rm vdW}$ the drastic increase of the {\em local} classical velocity
$\hbar k_{L}(r)$ suppress the probability for two particles to exist between $r$ and $r+dr$, 
which is proportional to $[m dr/\hbar k_{L}(r)]$. 
The suppression of the probability to find two particles at short distances can be seen in the WKB approximation to the two-body wave function:
\begin{equation}
\psi(r<r_{\rm vdW})=\frac{C}{k_{L}^{1/2}(r)}\sin\left[\int^{r}k_{L}(r')dr'+\frac{\pi}{4}\right],\label{psiWKB}
\end{equation}
where $C$ is a normalization constant and $k_{L}^2(r)=2\mu_{\rm 2b}[E_{\rm 2b}-v(r)]$. 
Therefore, the suppression of the wave function inside the potential well 
is simply related to factor $k_{L}(r)^{-1/2}$ which leads to amplitude suppression of 
the WKB wave function [Eq.~(\ref{psiWKB})] between $r$ and $r+dr$. [Fig.~\ref{Fig1}~(a)illustrates this effect by comparing
the fully numerical two-body wavefunction and the $k_{L}(r)^{-1/2}$ term from the WKB solution above.] Such suppression on the two-body 
level is directly reflected in the three-body system as the suppression of the probability to
find all three particles at short distances, thus eliminating (almost
fully) the necessity for accounting for the details of the interactions there. 
This suppression is manifested through the presence of a universal repulsive three-body barrier, resulting from the increase of the kinetic energy at short 
distances, for when particles approach distances smaller than $2r_{\rm vdW}$ [see Fig.~\ref{Fig1}~(b)]. Our results 
correctly reproduce the experimentally observed universality \cite{S_133Cs_IBK_1}-\cite{S_85Rb_JILA} within 15\%, and we
O expect them to be valid for broad Feshbach resonances. For narrow Feshbach resonances, however, other factors 
\cite{S_Schimdt,S_Jensen,S_Petrov,S_Gogolin,S_Wang}
could lead to deviations from such predictions, but this issue is only in the early stages of being investigated and quantified thus far.

\begin{figure}[htbp]
\includegraphics[width=0.9\columnwidth,angle=0,clip=true]{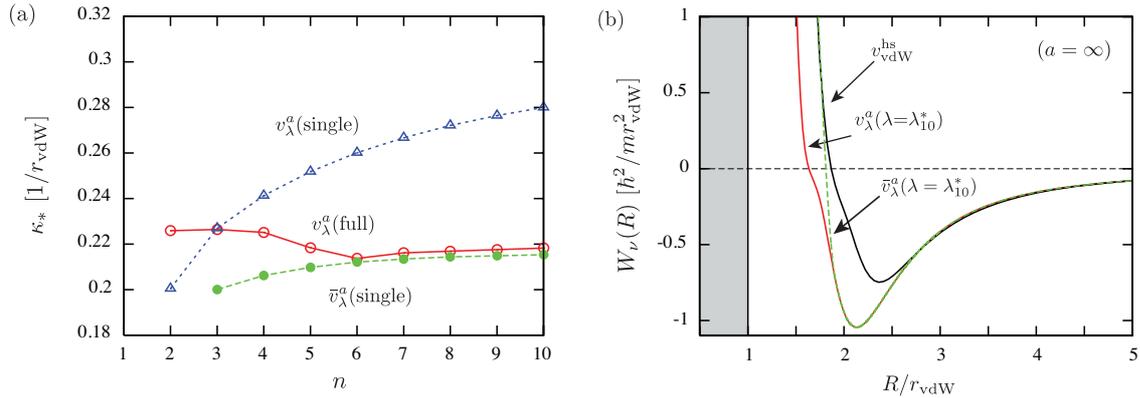}
\caption{
(a) Comparison between the three-body parameter $\kappa_*$
obtained from a single channel approximation (open triangles) as well as for our full numerical results (open circles)
using the $v_{\lambda}^a$ model. The single channel approximation can be improved by imposing a simple change in the adiabatic 
potentials near the barrier, as is shown in (b). There we smoothly connect the three-body potential for $v_{\lambda}^a$ (red
solid line) to the barrier obtained for the model potential $v_{\rm vdW}^{\rm hs}(r)$ [$+\infty$ for $r<0.8828r_{\rm vdW}$ and $-C_{6}/r^6$ otherwise]
(black solid line), resulting  in the potential labeled by $\bar{v}_{\lambda}^a$ (green solid line). This new potential is actually more
repulsive and has energies [filled circles in (a)] that are much closer to our full numerical calculations. 
}
\label{Fig2}
\end{figure}

This physical picture is, in fact, closely related to the strong three-body multichannel nature of the problem.
The increase of the local velocity $k_{L}(r)$, evidently, translates to the increase of the kinetic energy
in the problem. At the three-body level, this implies large values for the nonadiabatic
couplings $W_{\nu\nu'}(R)$, which in fact originate in the hyperradial kinetic energy. Therefore, the approximation of neglecting $W_{\nu\nu'}(R)$ in solving Eq.~(\ref{Sradeq}) can lead to a poor description of the system. 
In fact, the strong multichannel nature of the problem can be  illustrated by comparing the results obtained from a single channel approximation to
Eq.~(\ref{Sradeq}), i.e., $W_{\nu\nu'}(R)=0$ ($\nu\ne\nu'$), with our solutions of the fully coupled system of equations.
Figure~\ref{Fig2}~(a) shows the three-body parameter $\kappa_{*}$ [related to the energy of the lowest Efimov
state through the relation $\kappa_*=(m E/\hbar^2)^{1/2}$] obtained for the 
$v_{\lambda}^a$ model within the single channel approximation (open triangles) 
as well as our full numerical results (open circles). The disagreement between 
these quantities increases with the number of $s$-wave bound states $n$, meaning that 
the physics controlling the results becomes more and more multichannel in nature. 
Nevertheless, we find that by imposing a simple change in the adiabatic potentials near 
the barrier --- to make the barrier {\em more} repulsive [see Fig.~\ref{Fig2}~(b)] ---
the single channel approximation for $\kappa_{*}$ [filled circles in Fig.~\ref{Fig2}~(a)] 
reproduces the full numerical calculations much better. This agreement indicates that most of the 
nonadiabaticity of the problem is related to the exact shape of the barrier and that, to some extent, 
the effect of the nonadiabatic couplings is to make the effective potential $W_{\nu}$ more repulsive.
For these reasons and, of course, the universality of our full calculations, we believe that, despite the strong multichannel nature, 
the short-range barrier in the three-body effective potentials indeed offers a physically valid explanation of the approximate
universality of the three-body parameter for atomic systems whose long-range interactions are controlled by the van der Waals potential. 

In summary, we have elaborated our physical picture for the universality of the 3BP in terms
of the increase of the local classical velocity and demonstrated the strong multichannel nature
of the problem. This fact emphasizes the necessity to fully describe the three-body problem for additional realistic 
systems. In this vein, we have also recently shown that the universal principles analyzed here also apply to 
heteronuclear systems \cite{S_3BP_BBX} and to systems with small scattering lengths \cite{S_Dwave}. 
These results once more show that ultracold quantum gases offer a unique opportunity to study some of the 
most deep and fundamental aspects of few-body systems.

\begin{acknowledgements}
This work was supported by the U.S. National 
Science Foundation and by an AFOSR-MURI grant.
\end{acknowledgements}

\end{document}